# Some use of new wave-wave interaction conditions

## V.A. Buts

*National Science Centre "Kharkov Institute of Physics and Technology", Institute of Radio Astronomy of NAS of Ukraine, Kharkov, Ukraine*

## ABSTRACT

New conditions for wave-wave interaction are considered. It is shown that using these conditions allows us to discover and describe new features of wave-wave interaction. Specifically, it is shown that an electromagnetic wave in a stationary periodic medium can excite a wave with a different frequency. It is shown that waves with different frequencies can effectively interact in such a medium. It is shown that these conditions can reveal new features of three-wave interaction in nonlinear media. The use of new conditions in quantum mechanics has revealed a number of new features in the dynamics of particles and their interactions with each other. In particular, new energy levels have been discovered that can exist even in stationary periodic potentials. An assessment is given of the level of additive and multiplicative noise that can suppress the interaction process.

## 1. Introduction

Wave processes are the most common in nature. Among these processes, wave interactions play a significant role. Thus, in plasma physics, two fundamental physical processes are distinguished: wave-particle interactions and wave-wave interactions. The main results of the theory of wave interactions are based on local interaction conditions. This means that phase-matching conditions are imposed along each of the four axes of four-dimensional spacetime $(t, \vec{r})$:

$$\Delta \vec{k} = \sum_i \vec{k}_i = 0, \quad \Delta \omega = \sum_i \alpha_i \omega_i = 0; \quad \alpha_i = \pm 1 \qquad (1)$$

Here $\omega_i$ are the frequencies of the interacting waves, and $\vec{k}_i$ are the wave vectors of these waves. These conditions coincide with the conditions for resonant particle interaction. These conditions are often referred to as the law of conservation of momentum and the law of conservation of energy. Indeed, if we multiply the left-hand and right-hand sides of these conditions by Planck's constant, they truly describe the conservation laws of elementary processes during particle interactions.

Moreover, these are interactions in quantum mechanics. They are not applicable to conservation laws within classical electrodynamics. They only indicate the necessary phase relationships between interacting waves. Moreover, such a beautiful analogy hinders attempts to find other conditions for effective wave interaction. Of course, conditions (1) correctly describe both particle interactions and wave interactions. They underlie all wave-wave interaction processes in plasma physics. See, for example, in monographs and reviews [1-5]).

It should be kept in mind, however, that the processes of particle and wave interaction differ significantly. Particle interaction is indeed local, occurring within a small spatiotemporal region. Wave interaction most often occurs in a region whose spatiotemporal characteristics are significantly greater than the wavelength and wave period. It may turn out that a phase mismatch, for example along the time axis ($\Delta\omega \neq 0$), can be compensated for by a mismatch along one of the spatial axes ($\Delta k \neq 0$). Analysis of these features showed that such compensation is indeed possible. As a result, the wave interaction conditions took the form $\Delta\omega = (\Delta k \cdot V)$. The first considerations regarding these wave interaction features appeared in [6]. In [7] described some consequences of the new interaction conditions. It should be noted that there are some other results on generalizing the conditions of wave interaction under various conditions (see, for example, [8-10]). These generalized conditions apply to specific cases, for example, those where geometric optics methods can be used [8,10]. However, in all cases, these generalizations are based on considerations related to conditions (1).

In this paper, we continue to explore some of the consequences of using the new wave interaction conditions [6.7]. In the next (second) section, the problem and all the constraints used will be formulated. The basic system of truncated partial differential equations will be written out. In the third section, it will be shown that the use of the new conditions allows us to prove the fact that a wave propagating in a stationary, periodically inhomogeneous medium can excite another wave. Moreover, the frequency of the excited wave can be differed from the frequency of the excited wave. This result undermines the well-known notion (the well-known paradigm) that a wave in a stationary medium cannot excite waves with different frequencies. In other words, only elastic scattering processes can occur in a stationary medium. It is further shown that such waves (waves with different frequencies) can effectively exchange energy among themselves.

## 2. Problem Statement and Basic Equations

Let's consider the simplest model of wave interaction, in which new conditions for wave interaction can be realized. This is a three-wave interaction model, in which one of the interacting waves is a fixed wave. The characteristics of this wave do not change. The permittivity of the medium can be considered as such a wave. For example, consider a medium whose permittivity can be represented as:

$$\varepsilon = \varepsilon_0 + \tilde{\varepsilon}, \quad \tilde{\varepsilon} = q\cos(\vec{\kappa}\vec{r} - \Omega t), \quad q \ll 1 \tag{2}$$

Let's assume that two electromagnetic waves with different frequencies propagate in such a medium. We will be interested in the conditions for the effective interaction of these waves in such a medium. The equations for each of these waves are Maxwell's equations. From Maxwell's equations, it is easy to find an equation for the electric field vectors of the waves in such a medium:

$$\Delta\vec{E} - \frac{1}{c^2}\frac{\partial^2(\varepsilon\vec{E})}{\partial t^2} = -\vec{\nabla}\left(\frac{1}{\varepsilon}\vec{E}\cdot\vec{\nabla}\varepsilon\right) \tag{3}$$

Let all three vectors $(\vec{k}_0, \vec{k}_1, \vec{\kappa})$ lie in the plane of the page. Then equation (3) will take on a simpler form:

$$\Delta\vec{E} - \frac{1}{c^2}\frac{\partial^2(\varepsilon\vec{E})}{\partial t^2} = 0 \tag{4}$$

By assumption, there are two waves, therefore we will seek the solution to equation (4) as the sum of two terms:

$$\vec{E} = \vec{A}_0(\vec{r},t)\exp(-i\omega_0 t + i\vec{k}_0\vec{r}) + \vec{A}_1(\vec{r},t)\exp(-i\omega_1 t + i\vec{k}_1\vec{r}) \tag{5}$$

Here $k_0^2 = \omega_0^2\varepsilon_0/c^2$, $k_1^2 = \omega_1^2\varepsilon_0/c^2$.

Each term in (5) describes a plane electromagnetic wave, the amplitudes of which are functions of time and coordinates. Since the spatiotemporal inhomogeneity of the medium is assumed to be small, we will assume these amplitudes to be slowly varying functions of space and time.

$$\frac{\partial^2 A_j}{\partial z^2} \sim \frac{\partial^2 A_j}{\partial t^2} \sim q^2 \ll 1 \tag{6}$$

Let also $\vec{A}_0 = A_{0y} \equiv A_0$, and $\vec{A}_1 = A_{1y} \equiv A_1$; $\vec{k}_0 = \{k_{0x}, 0, k_{0z}\}$; $\vec{k}_1 = \{k_{1x}, 0, k_{1z}\}$; $\vec{\kappa} = \{\kappa_x, 0, \kappa_z\}$

$$k_0^2 = \omega_0^2 \varepsilon_0 / c^2, \quad k_1^2 = \omega_1^2 \varepsilon_0 / c^2 \quad V_j \equiv v_{phj} = \frac{\omega_j}{k_j}$$

Considering all these features and limitations, we can write the following system of equations to find slowly changing functions $A_j$:

$$\hat{L}_0 A_0 \equiv \left[k_{0z}\frac{\partial}{\partial z} + k_{0x}\frac{\partial}{\partial x} + \frac{\varepsilon_0 \cdot \omega_0}{c^2}\frac{\partial}{\partial t}\right] A_0 = -\frac{q}{4i}\frac{(\omega_1 \pm \Omega)^2}{c^2} A_1 \cdot \exp[+i \cdot \delta(\vec{r},t)] \tag{7}$$

$$\hat{L}_1 A_1 \equiv \left[k_{1z}\frac{\partial}{\partial z} + k_{1x}\frac{\partial}{\partial x} + \frac{\varepsilon_0 \cdot \omega_1}{c^2}\frac{\partial}{\partial t}\right] A_1 = -\frac{q}{4i}\frac{(\omega_0 \pm \Omega)^2}{c^2} A_0 \cdot \exp[-i \cdot \delta(\vec{r},t)] \tag{8}$$

Here $\delta(\vec{r},t) \equiv \Delta\vec{k} \cdot \vec{r} - \Delta\omega \cdot t$, $\Delta\vec{k} \equiv (\vec{k}_1 - \vec{k}_0 \pm \vec{\kappa})$, $\Delta\omega \equiv \omega_1 - \omega_0 \pm \Omega$

### 3. Excitation of a wave with frequency $\omega_1$ by a wave with frequency $\omega_0$ in a stationary medium ($\Omega = 0$).

Let $\hat{L}_0 \exp(i\delta) = 0$. This means:

$$\Delta\omega = (\Delta\vec{k} \cdot \vec{V}_0) \tag{9}$$

Let's use this property of the operator $\hat{L}_0$. We'll apply it to equation (8). Then equation (8) for determining the amplitude $A_1$ will take the form

$$\hat{L}_0 \hat{L}_1 A_1 = -\frac{q^2}{16 \cdot c^4}\left[(\omega_0 \pm \Omega)(\omega_1 \pm \Omega)\right]^2 A_1 \tag{10}$$

First, let's consider the case of a stationary medium ($\Omega = 0$). Furthermore, we will simplify the equations (where possible) in the future. For our purposes, it is sufficient to consider spatially one-dimensional dynamics. Equation (10) can then be rewritten as

$$\left[\frac{\partial^2}{\partial z^2} + \frac{1}{V_0 V_1}\frac{\partial^2}{\partial t^2} + \frac{\partial}{\partial z}\frac{\partial}{\partial t}\left(\frac{1}{V_0} + \frac{1}{V_1}\right)\right] A_1 = -\frac{q^2 k_0 k_1}{16} A_1 \tag{11}$$

To evaluate the characteristic spatial and temporal quantities that determine the process of excitation of a wave with frequency, we determine the temporal dynamics at fixed coordinates ($z = const$):

$$\frac{\partial^2 A_1}{\partial t^2} + \frac{q^2 \omega_0 \omega_1}{16} A_1 = 0 \tag{12}$$

The amplitude of the excited wave periodically varies with a characteristic frequency $\Omega_{ch} = (q/4)\sqrt{\omega_0 \omega_1}$. The dispersion diagram of these waves is shown in Figure 1. Similarly, the characteristic scale of spatial variation of this amplitude ($t = const$) can be estimated:

$$\frac{\partial^2 A_1}{\partial z^2} + \frac{q^2 k_0 k_1}{16} A_1 = 0 \qquad (13)$$

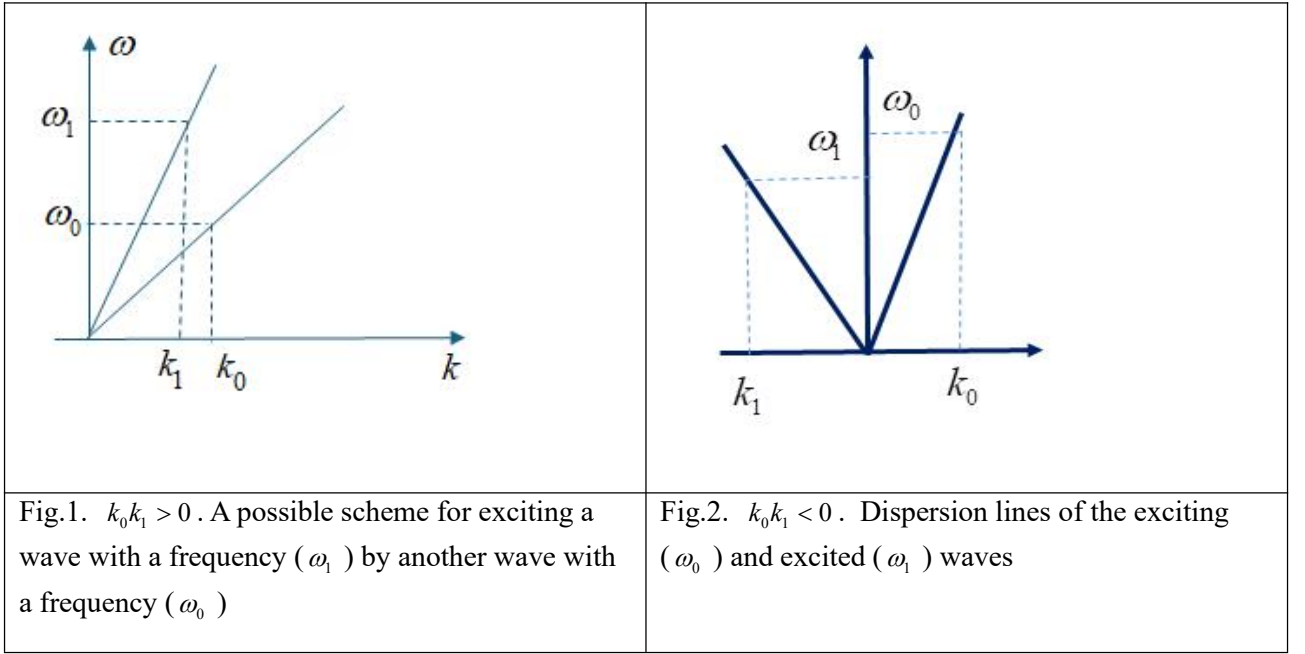

Fig.1. $k_0 k_1 > 0$. A possible scheme for exciting a wave with a frequency ($\omega_1$) by another wave with a frequency ($\omega_0$)

Fig.2. $k_0 k_1 < 0$. Dispersion lines of the exciting ($\omega_0$) and excited ($\omega_1$) waves

Spatial "dynamics" are richer. If $k_0 k_1 > 0$, then the amplitude of the excited wave periodically changes with a spatial period equal to $K_{ch} = (q/4)\sqrt{k_0 k_1}$. If $k_0 k_1 < 0$, then the spatial dynamics resemble Bragg reflection, with the only difference being $\omega_1 \neq \omega_0$ (see Fig. 2). The change in amplitude $A_0$ can be found by substituting the solution of equation (11) into equation (7). However, restricting the process under consideration to the excitation of a wave by the field of another wave is sufficiently limited by the fact that the average value $A_0$ does not change ($\langle A_0 \rangle = const$). Using (9), we obtain the following expression for the frequency of the excited wave

$$\omega_1 = \pm \frac{\kappa V_0 V_1}{(V_1 - V_0)}, \quad V_1 \neq V_0 \qquad (14)$$

### 4. Interaction of Waves with Different Frequencies

Above, we examined the process of waves generating new waves in stationary media. In this section, we will demonstrate that it is possible to study the interaction of such waves as well.

To do this, we will rewrite the system of equations (7), (8):

$$\left[ \frac{c}{\sqrt{\varepsilon_0}} \frac{\partial A_0}{\partial z} + \frac{\partial A_0}{\partial t} \right] = -\alpha_{10} A_1 \cdot \exp\left[ +i \cdot \delta_+(z,t) \right] \qquad (15)$$

$$\left[ \frac{c}{\sqrt{\varepsilon_1}} \frac{\partial A_1}{\partial z} + \frac{\partial A_1}{\partial t} \right] = -\alpha_{01} A_0 \cdot \exp\left[ -i \cdot \delta_-(z,t) \right] \qquad (16)$$

Here $\alpha_{10} = \frac{q}{4i} \frac{(\omega_1 + \Omega)^2}{\varepsilon_0 \omega_0}$, $\alpha_{01} = \frac{q}{4i} \frac{(\omega_0 - \Omega)^2}{\varepsilon_1 \omega_1}$, $\delta_\pm(\vec{r},t) \equiv \Delta \vec{k}_\pm \cdot \vec{r} - \Delta \omega_\pm \cdot t$, $\Delta \vec{k}_\pm \equiv (\vec{k}_1 - \vec{k}_0 \pm \vec{\kappa})$,

$\Delta \omega_\pm \equiv \omega_1 - \omega_0 \pm \Omega$

In deriving this system of equations, we limited ourselves to a one-dimensional inhomogeneity. Furthermore, we considered that one wave can interact with one component of the inhomogeneity $\delta_+(\vec{r},t)$ and another wave can interact with another ($\delta_-(\vec{r},t)$). Let us use the characteristics of equations (15) and (16):

$$\frac{dt}{1} = \frac{dz}{V_0} = -\frac{dA_0}{\alpha_{10} A_1 exp(i\delta_+)}, \qquad \frac{dt}{1} = \frac{dz}{V_1} = -\frac{dA_1}{\alpha_{01} A_0 exp(-i\delta_-)} \qquad (17)$$

These characteristics allow us to replace the two partial differential equations (15) and (16) with four ordinary differential equations:

$$\frac{dA_0}{dt} = -\alpha_{10} A_1 \exp(i\delta_+), \quad \frac{dz}{dt} = V_0 \; ; \quad \frac{dA_1}{dt} = -\alpha_{01} A_0 \exp(-i\delta_-), \quad \frac{dz}{dt} = V_1 \; ; \qquad (18)$$

Using the solutions of the second and fourth equations ($z = V_0 t$, $z = V_1 t$) we obtain the following expressions for $\delta_\pm(\vec{r},t)$:

$$\delta_+ = (V_0 \cdot \Delta k_+ - \Delta\omega_+) \cdot t \; , \; \delta_- = (V_1 \cdot \Delta k_- - \Delta\omega_-) \cdot t \qquad (19)$$

We can impose the following conditions on these quantities

$$\delta_+ = \delta_- = 0 \; ; \; \Delta\omega_+ = V_0 \Delta k_+ \; , \; \Delta\omega_- = V_1 \Delta k_- \qquad (20)$$

We are considering in the most interesting case: $\Omega = 0$. Then formula (20) gives the following expression for the frequency of the excited wave

$$\omega_1 = \omega_0 \frac{(V_0 + V_1)}{(3V_0 - V_1)} \qquad (21)$$

Using the system of equations (18) and conditions (20), the following useful relationships can be obtained

$$\frac{d^2 A_j}{dt^2} = -\Omega_q^2 A_j \qquad j = \{0,1\} \qquad A_0^2 - \mu A_1^2 = const \qquad (22)$$

Where $\Omega_q^2 = \dfrac{q^2 (\omega_1 + \Omega)^2 (\omega_0 - \Omega)^2}{16 \cdot \varepsilon_1 \cdot \varepsilon_0 \cdot \omega_1 \cdot \omega_0}$, $\mu = \dfrac{(\varepsilon_1 \cdot \omega_1)}{(\varepsilon_0 \cdot \omega_0)} \left(\dfrac{\omega_1 + \Omega}{\omega_0 - \Omega}\right)^2$

From the first equation, we can immediately determine the characteristic time of energy exchange between interacting waves ($T_q \sim (25/q\omega)$). The second relation indicates that the exchange process is bounded by a certain integral. Note that, as follows from (18), the complex amplitudes $A_0$ and $A_1$ are shifted relative to each other by $\pi/2$.

### 5. Wave-wave Interaction in Nonlinear Media.

The above-described conditions for wave synchronism in inhomogeneous media are apparently the easiest to implement and use for experimental verification of the formulated interaction conditions. However, the formulated conditions can also be implemented in many other physical processes. Such processes can include wave interactions in nonlinear media. In this section, we consider three-wave interactions in nonlinear media. The main purpose of this consideration is to demonstrate the fact that new conditions of interaction can reveal new features of many other processes.

The equations that describe the interaction of three waves ($\omega_0, \omega_1, \omega_2$ ; $\vec{k}_0, \vec{k}_1, \vec{k}_2$) in a nonlinear medium can be represented as (see, for example, [1]):

$$\frac{\partial a_0}{\partial l_0} \equiv \dot{a}_0 + (\vec{V}_0 \vec{\nabla}) a_0 = -\sigma_0 a_1 a_2 \exp(-i\delta) \equiv f_0$$

$$\frac{\partial a_1}{\partial l_1} \equiv \dot{a}_1 + (\vec{V}_1 \vec{\nabla}) a_1 = \sigma_1 a_0 a_2^* \exp(i\delta) \equiv f_1 \qquad (23)$$

$$\frac{\partial a_2}{\partial l_2} \equiv \dot{a}_2 + (\vec{V}_2 \vec{\nabla}) a_2 = \sigma_2 a_0 a_1^* \exp(i\delta) \equiv f_2$$

Here, the dots denote partial derivatives with respect to time. $\delta(\vec{r},t) \equiv \Delta \vec{k} \cdot \vec{r} - \Delta \omega \cdot t$ is the detuning, which we do not consider a slow function of coordinates and time; $\Delta \omega = \omega_0 - \omega_1 - \omega_2$ ; $\Delta \vec{k} = (\vec{k}_0 - \vec{k}_1 - \vec{k}_2)$ ; $\vec{V}_i$ are the group velocities of the waves; $\sigma_i$ are the matrix elements of the nonlinear interaction.

The left-hand side of each equation in system (23) is represented as derivatives along the characteristic directions. For clarity, we will consider the interaction of waves with positive energy ( $\sigma_i > 0$ ), and we will also focus on decay processes. The equations for the characteristics of each equation in system (23) can be written as:

$$\frac{dt}{1} = \frac{dx}{V_{ix}} = \frac{dy}{V_{iy}} = \frac{dz}{V_{iz}} = \frac{da_i}{f_i} \qquad (24)$$

Unit vectors directed along these characteristics will have the form: $\vec{l}_i = \{V_{ix}, V_{iy}, V_{iz}, 1\}/N_i$.

Here $N_i = \sqrt{V_{ix}^2 + V_{iy}^2 + V_{iz}^2 + 1}$ . The condition for synchronism of the interacting waves will be the condition of parallelism of the characteristic lines of the hyperplane $\delta(\vec{r},t) \equiv \Delta \vec{k} \cdot \vec{r} - \Delta \omega \cdot t = const$ , i.e., the condition:

$$\Delta \omega - \Delta \vec{k} \cdot \vec{V}_i = 0 \qquad (25)$$

The linear stage of the decay process proceeds as instability. At this stage, the wave with the maximum frequency can be considered fixed ( $a_0 = const$ ). In this case, system (23) can be conveniently rewritten as:

$$\frac{\partial a_1}{\partial l_1} = \sigma_1 a_0 a_2^* \exp(i\delta) \qquad \frac{\partial a_2^*}{\partial l_2} = \sigma_2 a_1 a_0^* \exp(-i\delta) \qquad (26)$$

It is convenient to rewrite system (26) as a single second-order partial differential equation with constant coefficients:

$$\frac{\partial^2 a_1}{\partial l_1 \partial l_2} = \sigma_1 \sigma_2 |a_0|^2 a_1 \qquad (27)$$

In deriving (27), we used the fact that the derivatives along the characteristic directions of the function $\delta(\vec{r},t)$ are equal to zero. Substituting the solution in the form $\sim \exp(i\Omega t - i\vec{\kappa}\vec{r})$ into (27), we obtain the following dispersion equation, which determines the relationship between the frequency $\Omega$ and the vector $\vec{\kappa}$ :

$$\Omega^2 - \Omega \cdot \vec{k}(\vec{V}_1 + \vec{V}_2) + \left[ k^2(\vec{V}_1 \vec{V}_2) + \sigma_1 \sigma_2 |a_0|^2 \right] = 0 \qquad (28)$$

From the synchronism conditions (25) it follows that synchronism will be possible and under the condition of equality of group velocities ( $\vec{V}_1 = \vec{V}_2$ ). Solving equation (28) for $\Omega$ , in this case we obtain:

$$\Omega = \vec{k}\vec{V} \pm i|a_0|\sqrt{\sigma_1 \sigma_2} \qquad (29)$$

The imaginary part of the frequency $\left( \text{Im}\,\Omega = |a_0|\sqrt{\sigma_1 \sigma_2} \right)$ determines the increment of decay instability. It is of interest to discover new possibilities for decays that do not fit within the known decay conditions. Let us consider the simplest. Let us assume that a transverse wave decays into a transverse wave and one of the eigenmodes of a magnetized plasma waveguide ( $t_0 \to t_1 + l$ ). We will assume that all three waves fit into the linear dispersion region (see Fig.3). In this case, the group velocities of all three waves coincide ($\vec{V}_0 = \vec{V}_1 = \vec{V}_2 = \vec{V}$). Moreover, the group velocities in this case coincide with the phase velocities. The synchronism conditions (25) in this case will be satisfied for any triplet of waves. Indeed, conditions (25) in this case take the form of an identity:

$$\omega_0 - \omega_1 - \omega_2 = \left(\frac{\omega_0}{V} - \frac{\omega_1}{V} - \frac{\omega_2}{V}\right) V \quad (30)$$

Thus, in the case under consideration, the decay process may involve many wave triplets. Moreover, these waves may differ little in their characteristics. In this case, as shown in [11], the decay process becomes chaotic. It should be noted that the case under consideration of the coincidence of all velocities can be solved analytically not only at the linear stage but also at the nonlinear stage. For this, it is sufficient to use the effective potential method (see, for example, [1]). Such a solution is beyond the scope of our interests. Moreover, because, as we noted above, the decay process will be chaotic, this solution loses its meaning.

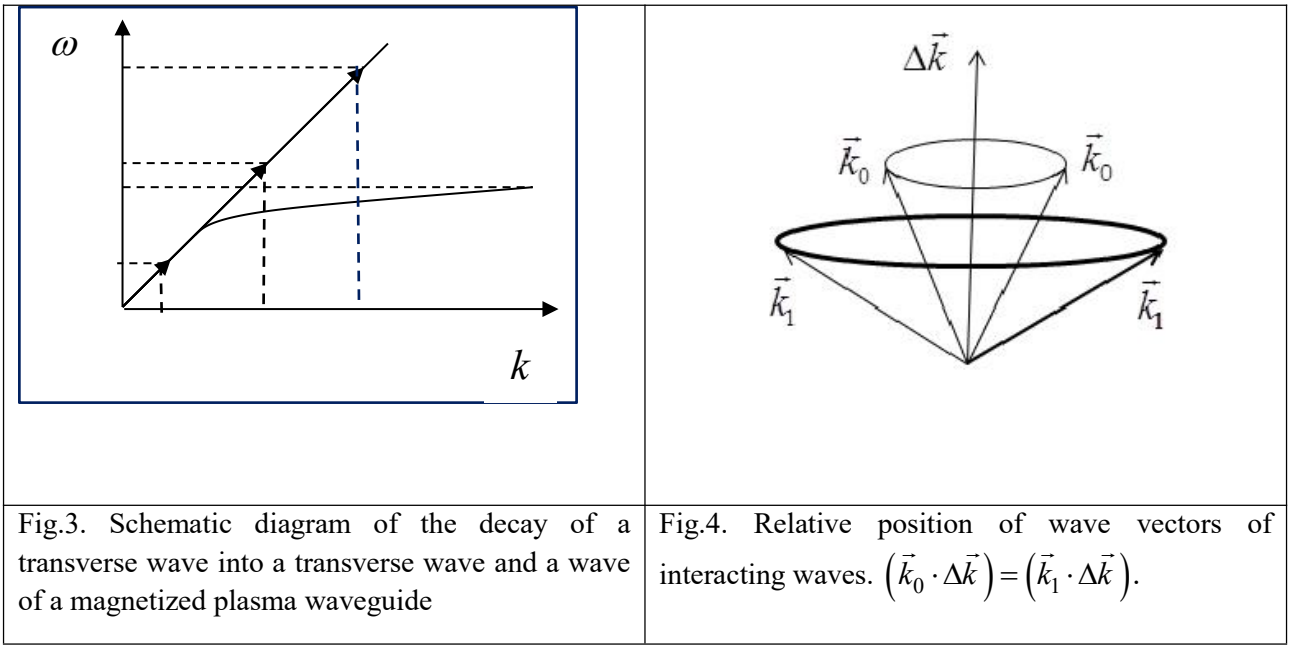

| Fig.3. Schematic diagram of the decay of a transverse wave into a transverse wave and a wave of a magnetized plasma waveguide | Fig.4. Relative position of wave vectors of interacting waves. $(\vec{k}_0 \cdot \Delta \vec{k}) = (\vec{k}_1 \cdot \Delta \vec{k})$. |
|---|---|

### 6. Interaction of waves in quantum systems

Interactions in quantum systems are wave-like. Therefore, it is natural to expect that interactions in such systems will also exhibit, in addition to the well-known synchronicity conditions ($\Delta \omega = 0$ and $\Delta \vec{k} = 0$), generalized conditions similar to conditions (9). We will demonstrate that such processes are indeed realized in quantum systems. To do this, we consider the dynamics of a quantum system moving in a potential that is periodic in time and space: $U = U_0 + q \cdot \cos(\Omega \cdot t - \vec{\kappa} \cdot \vec{r})$. We will assume that the periodic perturbation is small ($q \ll 1$). The wave function describing the quantum system satisfies the Schrödinger equation:

$$i\hbar \frac{\partial \psi}{\partial t} + \frac{\hbar^2}{2m} \Delta \psi + e \cdot U \cdot \psi = 0 \quad (31)$$

We will seek a solution to equation (31) as the sum of two functions: $\psi(\vec{r},t) = \psi_0(\vec{r},t) + \psi_1(\vec{r},t)$, Each of these terms represents a solution to the unperturbed Schrödinger equation. Each term represents a wave (a probability wave). In the absence of a perturbation, these waves are independent. In the presence of a periodic perturbation of the potential, these waves begin to interact. The complex amplitudes of these waves will change in space and time. Finding the conditions for the effective interaction of these waves will be our task. Thus, we will seek a solution in the form:

$$\psi(\vec{r},t) = \sum_{i=0}^{1} A_i(\vec{r},t) \cdot \exp\left(i \cdot \omega_i t - i \cdot \vec{k} \cdot \vec{r}\right). \tag{32}$$

In what follows, when discussing a quantum system, we will be discussing the dynamics of a charged particle. Then, the particle's energy in each of the states under consideration is $E_i = \hbar \omega_i = k_i^2 \cdot \hbar^2 / 2m$, and its momentum is $\vec{p}_i = \hbar \vec{k}_i$. The wave vectors and frequencies in (32) are related by the relation: $\omega_i = k_i^2 \hbar / 2m$. The wave vector is $\vec{k}_i = m\vec{V}_i / \hbar$. Substituting (32) into the Schrödinger equation, we can obtain the following system of differential equations for determining slowly varying amplitudes $A_i$:

$$\frac{\partial A_0}{\partial l_0} \equiv \frac{\partial A_0}{\partial t} - \frac{\hbar}{m}\left(\vec{k}_0 \vec{\nabla}\right) A_0 = -\frac{e \cdot q}{i \cdot \hbar} A_1 \cdot \exp(i\delta)$$

$$\frac{\partial A_1}{\partial l_1} \equiv \frac{\partial A_1}{\partial t} - \frac{\hbar}{m}\left(\vec{k}_1 \vec{\nabla}\right) A_1 = -\frac{e \cdot q}{i \cdot \hbar} A_0 \cdot \exp(-i\delta). \tag{33}$$

Here, as in the scattering of electromagnetic waves, $\delta = \Delta\omega \cdot t - \Delta\vec{k} \cdot \vec{r}$; $\Delta\omega = \omega_1 - \omega_0 \pm \Omega$; $\Delta\vec{k} = \vec{k}_1 - \vec{k}_0 \pm \vec{\kappa}$. $\hat{L}_0 \delta = \Delta\omega - \vec{V}_0 \Delta\vec{k} = 0 \quad \hat{L}_1 \delta = \Delta\omega - \vec{V}_1 \Delta\vec{k} = 0$

The characteristic equations of the homogeneous equations of system (33) have the form:

$$\frac{dt}{1} = -\frac{dx \cdot m}{k_{ix} \hbar} = -\frac{dy \cdot m}{k_{iy} \hbar} = -\frac{dz \cdot m}{k_{iz} \hbar} \tag{34}$$

The condition of synchronism will be the condition that these characteristic lines will be parallel to the hyperplane: $\delta(\vec{r},t) = const$:

$$\Delta\omega - \frac{\hbar}{m}\left(\vec{k}_0 \Delta\vec{k}\right) = 0; \quad \Delta\omega - \frac{\hbar}{m}\left(\vec{k}_1 \Delta\vec{k}\right) = 0 \ . \tag{35}$$

From (35), in addition to the well-known conditions of synchronism $\Delta\omega = 0$ and $\Delta\vec{k} = 0$, new conditions follow: $\left(\vec{k}_0 \cdot \Delta\vec{k}\right) = \left(\vec{k}_1 \cdot \Delta\vec{k}\right)$, $\Delta\vec{k} \neq 0$ see Fig. 4. From this condition, in particular, it follows that when particles with energy $\hbar\omega_0$ and momentum $\hbar\vec{k}_0$ move in a potential that periodically changes in space and time, they can be transformed into particles with energies:

$$\hbar\omega_1 = \hbar\omega_0 \mp \hbar\Omega + \hbar^2 \left(\vec{k} \cdot \Delta\vec{k}\right)/m . \tag{36}$$

If the usual synchronism conditions ($\Delta\vec{k} = 0$) are met, we obtain the well-known result regarding the appearance of particles with quasi-energies. It is the third term that distinguishes the obtained result from the known one. The question arises about the time interval (or distance) during which such a transformation will occur. For this, it is convenient to move from the system of equations (33) to a single second-order equation:

$$\frac{\partial^2 A_0}{\partial l_0 \partial l_1} = -\frac{e^2 q^2}{\hbar^2} A_0 . \tag{37}$$

In deriving (37), we took into account, as in the previous sections, the fact that the derivatives of the function $\delta(\vec{r},t)$ along the characteristics are equal to zero. Equation (37) is an equation with constant coefficients. Substituting the solution $A_0 \sim \exp\left[i\left(\omega_h t - \vec{k}_h \vec{r}\right)\right]$ into it, we obtain a dispersion equation that relates $\omega_h$ and $\vec{k}_h$:

$$\omega_h^2 + \omega_h \cdot \left[\left(\vec{k}_0 \vec{k}_h\right) + \left(\vec{k}_1 \vec{k}_h\right)\right]\frac{\hbar}{m} + \left[\left(\vec{k}_0 \vec{k}_h\right)\left(\vec{k}_1 \vec{k}_h\right)\left(\frac{\hbar}{m}\right)^2 - \left(\frac{e \cdot q}{\hbar}\right)^2\right] = 0 . \tag{38}$$

At $\vec{k}_h \to 0$ from (38), in particular, it follows that $\omega_h = e \cdot q / \hbar$, i.e., over times of the order $\hbar/eq$ the particles with energies $\hbar\omega_0$ are transformed into particles with energies determined by formula (36).

With longer interactions, the exchange process will change sign, and the "new particles" will be transformed into the original ones. To find the necessary parameters and characteristic temporal and spatial characteristics of this exchange, it is easiest to use the considerations in Section 4. For this, system of equations (33) can be replaced by a system of eight ordinary differential equations:

$$\frac{dA_0}{dt} = \alpha A_1 \exp(i\delta_+) \quad \frac{dA_1}{dt} = \alpha A_0 \exp(-i\delta_-), \quad \frac{d\vec{r}_0}{dt} = -\frac{\hbar \vec{k}_0}{m}, \quad \frac{d\vec{r}_1}{dt} = -\frac{\hbar \vec{k}_1}{m}, \tag{39}$$

Where $\delta_+ = \left(\vec{V}_0 \cdot \Delta\vec{k}_+ - \Delta\omega_+\right)\cdot t$, $\delta_- = \left(\vec{V}_1 \cdot \Delta\vec{k}_- - \Delta\omega_-\right)\cdot t$, $\Delta\omega_\pm = \omega_1 - \omega_0 \pm \Omega$; $\Delta\vec{k}_\pm = \vec{k}_1 - \vec{k}_0 \pm \vec{\kappa}$,

$\alpha = ieq/\hbar$,

The last six of these equations have trivial solutions. We'll use these solutions. Additionally, we'll impose the following constraints: $\delta_+ = 0, \vec{V}_0 \cdot \Delta\vec{k}_+ - \Delta\omega_+ = 0;$ $\delta_- = 0, \vec{V}_0 \cdot \Delta\vec{k}_- - \Delta\omega_- = 0;$

In this case, system (33) can be reduced to two ordinary differential equations:

$$\frac{dA_0}{dt} = \alpha A_1, \quad \frac{dA_1}{dt} = \alpha A_0 \tag{40}$$

This system is equivalent to the equations of a linear pendulum:

$$\frac{d^2 A_j}{dt^2} + \left(\frac{eq}{\hbar}\right)^2 A_j = 0, \quad j = \{0,1\}: \tag{41}$$

The required solutions of these equations have the following integral

$$A_0^2 + A_1^2 = const \tag{42}$$

Similar equations can be derived to determine the characteristic spatial parameters of the interaction process. For example, to determine the characteristic parameter of amplitude changes along the z-axis, the following equations can be derived:

$$\frac{dA_0}{dz} = -\frac{ieq}{\hbar V_{0z}} A_1, \quad \frac{dA_1}{dz} = -\frac{ieq}{\hbar V_{1z}} A_0 \tag{43}$$

Depending on the sign of the velocity product, the spatial dynamics of the particles will vary. If this product is positive, the particles periodically exchange energy and propagate in the same direction. If the sign is negative, the interaction process is analogous to Bragg reflection. From formulas (43) and similar ones for other spatial directions, the following characteristic spatial parameter of energy exchange between particles follows:

$$L \approx \left(\hbar \sqrt{V_{0j} V_{1j}} / eq\right); \quad j = \{x, y, z\}, \tag{44}$$

It is useful to evaluate the influence of fluctuations on the interaction process. The following equations can be used to estimate this influence:

$$i\hbar \frac{\partial \psi}{\partial t} + \frac{\hbar^2}{2m} \Delta \psi + e \cdot U \cdot \psi = (e \cdot \xi), \quad i\hbar \frac{\partial \psi}{\partial t} + \frac{\hbar^2}{2m} \Delta \psi + e \cdot U \cdot \psi = (e \cdot \xi) \psi \tag{45}$$

Where $\xi(t)$ – fluctuation

The first equation of system (45) describes the influence of additive fluctuations. The second describes the influence of multiplicative fluctuations. It is possible to obtain solutions to the equations (45); however, their form is very cumbersome. Therefore, below, to estimate the magnitude of the influence of fluctuations on the interaction process, we will simplify the problem. First, consider the dynamics of a quantum system, , which is subject only to fluctuation forces. System of equations (45) then takes on the following simple form:

$$\frac{\partial \psi}{\partial t} = f(t), \quad \frac{\partial \psi}{\partial t} = f(t) \cdot \psi \tag{46}$$

Where $f(t) = \frac{ie}{\hbar} \xi(t)$

For simplicity, we will assume that the fluctuations are delta-correlated white noise:

$$\langle f(t) \rangle = 0, \quad \langle f(t) f(t_1) \rangle = D\delta(t - t_1) \tag{47}$$

By solving the first equation of system (46) one can obtain the well-known expression

$$\sqrt{\langle \psi(t_1) \cdot \psi(t_2) \rangle} = \sqrt{D \cdot (t_1 - t_2)} = \sqrt{D \cdot \tau}, \quad (48)$$

From this formula it follows that a change in the wave function by unity under the influence of additive fluctuations occurs over a time $T_{af} \approx (1/D)^2$. To solve the second equation of system (46), it is necessary to split the correlation ($\langle f(t)\psi \rangle$). For this, we will use the method of variational (functional) derivatives [12] and the relation that follows from this method (the Furutsu-Novikov formula)

$$\langle f(t) R[f(t)] \rangle = \int_t \langle f(t) f(\tau) \rangle \left\langle \frac{\delta R[f(\tau)]}{\delta f(\tau)} \right\rangle d\tau, \quad (49)$$

where $R[z]$ — arbitrary functional from $z$, $\delta R / \delta f$ - variational derivative.

Considering that $\delta \psi / \delta f = \psi$ and $\langle f(t)\psi \rangle = D \cdot \psi$ from the second equation of system (46) one can obtain the following equation for finding ($\langle \psi \rangle$):

$$\left\langle \frac{\partial \psi}{\partial t} \right\rangle = D \cdot \langle \psi \rangle \quad (50)$$

From this equation we find that the time of change of the average value of the wave function under the influence of multiplicative fluctuations is significantly less than the same change under the influence of additive fluctuations: ($T_{Mf} \approx (1/D), D \ll 1$). Comparing the times of change of wave functions in the presence of a regular periodic inhomogeneity (see (38)) we can calculate that if the inequality is satisfied

$$\frac{\hbar}{eq} \ll \frac{1}{D} \quad (51)$$

then the existing fluctuations will not significantly affect the process of particle interaction

## 7. Conclusion

Thus, the results obtained above demonstrate that the new wave interaction conditions are a useful tool for discovering new features of such an important process as wave-wave interaction. Let us briefly highlight the most important of these.

First, it's worth noting the result that a wave propagating in a periodic stationary medium can excite another wave. Importantly, the frequency of the excited wave can be differed from the frequency of the wave that excited it. This result challenges our notion (our paradigm) that only elastic processes can occur in a stationary medium.

It should be noted that a similar result was previously described in the paper [7]. However, the experimental conditions for observation were difficult. The result presented in section 3 can be easily observed. However, it should be noted that a system containing a wave is, strictly speaking, not stationary.

It is important that such waves can interact (section 4). Clearly, these results are primarily important for diagnostic purposes.

An important result is described in the fifth section. This result exemplifies the successful use of new conditions in studying wave interactions in nonlinear media. However, it is not only an example. The considered scheme for the stochastic decay of a transverse wave into a transverse wave and a Langmuir wave can be useful for plasma heating. This is especially true if the Langmuir wave is an Ion Langmuir wave (this wave interacts most effectively with the ion component of the plasma). In this case, the considered scheme offers the most direct way to deliver energy for plasma heating.

The most important results are obtained in the sixth quantum section. It is shown that even in a stationary periodic potential, particle energy changes. This result can be interpreted as the movement of particles from one hill to another. These same results can be considered as particle interactions. Moreover, this interaction applies not only to particles with the same energy (as in Bragg reflection), but also to particles with different energies. These results significantly enhance the specific features of particle dynamics in solids and crystals. Equally important is the assessment of the influence of the fluctuation level (50), which will not disrupt the particle interaction process.

**Acknowledgments**

The author is grateful to the Corresponding Member of the Academy of Sciences of Ukraine Dmitry Mikhailovich Vavriv for numerous fruitful discussions of the results